\documentclass[12pt]{article}
\usepackage[latin1]{inputenc}
\usepackage{empheq}
\usepackage{amsmath}
\usepackage{graphicx}
\usepackage{amssymb}
\usepackage{amsfonts}
\usepackage{latexsym}
\usepackage{float}
\usepackage{epstopdf}

\def\bea{\begin{equation}}
\def\eea{\end{equation}}
\def\beq{\begin{eqnarray}}
\def\eeq{\end{eqnarray}}

\def\ln{\,\mbox{ln}\,}

\def\Tr{\,\mbox{Tr}\,}

\def\Box{\square}

\def\al{\alpha}
\def\be{\beta}
\def\ch{\chi}
\def\ga{\gamma}\def\de{\delta}
\def\vp{\varepsilon}

\def\ka{\kappa}
\def\la{\lambda}
\def\na{\nabla}

\def\si{\sigma}
\def\om{\omega}
\def\ph{\varphi}

\def\th{\theta}

\def\Ga{\Gamma}
\def\De{\Delta}
\def\La{\Lambda}

\def\Om{\Omega}

\begin{document}

\title{Quantum Einstein-Cartan theory with the Holst term}

\author{
Ilya L. Shapiro$^{1}$\footnote{Also at Tomsk State Pedagogical University.
\quad Email address: shapiro@fisica.ufjf.br}
\ \  and \ \
Poliane M. Teixeira\footnote{
\quad Email address: poliane@fisica.ufjf.br}
\\ \\
Departamento de F\'{\i}sica, ICE, Universidade Federal de Juiz de Fora
\\
Campus Universit\'{a}rio - Juiz de Fora, MG, Brazil  36036-900
}

\date{}
\maketitle

\begin{quotation}
\noindent
{\large{\bf Abstract.}}
Holst term represents an interesting addition to the Einstein-Cartan
theory of gravity with torsion. When this term is present the contact
interactions between vector and axial vector fermion currents gain an
extra parity-violating component. We re-derive this interaction using
a simple representation for the Holst term. The same representation
serves as a useful basis for the calculation of one-loop divergences in
the theory with external fermionic currents and cosmological constant.
Furthermore, we explore the possibilities of the on-shell version of
renormalization group and construct the equations for the running of
dimensionless parameters related to currents and for the effective
Barbero-Immirzi parameter.

{\it MSC:} \ 81T15, 
             83D05  
          11.10.Gh, 
          04.50.Kd  
\vskip 4mm

PACS: $\,$
04.20.-q    
$\,\,$
98.80.-k    
\vskip 1mm

Keywords:  Barbero-Immirzi parameter, Einstein-Cartan theory,
Renormalization group.

\end{quotation}

\section{Introduction}

Einstein-Cartan theory attracts growing interest (see, e.g.,
\cite{freidel}, \cite{Fermion}, \cite{EC-phen} and references
therein) because it represents a simplest possible extension
of General Relativity (GR) related to the introduction of
torsion field. The presence of torsion enables one to enrich
the theory further by implementing the Holst term \cite{Holst},
which emerge naturally in the framework of loop quantum gravity
\cite{Ashtekar,PR,freidel,Alex}.
This parity-violating term should attract a special interest
since it can, in principle, yield some measurable observables for
detecting quantum gravity. In order to better understand this point
let us remember that in the Einstein-Cartan theory torsion becomes
relevant only in the presence of fermion currents. After being
integrated out, torsion provides contact
interactions between such currents. Obviously, the main
possibility for the Holst term, in this respect, is related to
the generation of parity-violating contact interaction between vector
and axial vector fermion currents. The first purpose of the present
communication is to present a very simple derivation of the Holst term
in terms of irreducible components of the torsion tensor. We show that
the new term is the simplest possible parity-violating scalar, and
hence the Barbero-Immirzi parameter \cite{barbero,immirzi} can be
seen as an extra non-minimal parity-violating extension of the
Einstein-Cartan action. Using this new form we recalculate the
contact interaction between fermion currents depending on
Barbero-Immirzi parameter and meet good correspondence with the
previous results of other authors \cite{mercuri,PR,freidel,Khriplovich}.

The main motivation for the Barbero-Immirzi parameter is related
to Quantum Gravity (QG), so it is natural to see what can be the
role of such term in the loop corrections. Since quantum GR and
also quantum Einstein-Cartan theory is not renormalizable,
this issue can not be addressed in the conventional framework
of perturbative quantum field theory for the metric and torsion.
The existing publications in this direction use very different
approaches. The first of them is based on the functional
renormalization group \cite{Reuter}. This powerful method is
essentially non-perturbative, and in case of QG there is no
perturbative limit, at least in the case of quantum GR. At the
same time, there is a known difficulty related to the gauge-fixing
dependence of the results of the functional renormalization group
applied to gauge theories \cite{Giess,Lavrov-FRG} (see also many
other references therein). In fact, the gauge-fixing dependence
in this theory persists on-shell \cite{Lavrov-FRG} and leads to
the gauge dependent $S$-matrix and possibly all other relevant
observables. One can expect that the same strong gauge dependence
will take place also in the case of QG, and this creates certain
difficulty for the physical interpretation of the results of this
approach.

Another possibility is to rely on the renormalization group
equations extracted from the quadratic one-loop divergences.
This is technically possible, however the ambiguities which one
usually meets in such a formulation are very strong and even go
beyond the gauge fixing ambiguities. This aspect of QG has
attracted significant interest recently, and the net result is
that these ambiguities are generally uncontrollable (see, e.g.,
\cite{Toms} and further references therein). For the Einstein-Cartan
theory with the Barbero-Immirzi parameter this scheme was applied
recently in \cite{Benedetti}, where the previous results for the
one-loop divergences in the quantum GR with interacting
fermion currents \cite{bavi-81} have been used.
\par
In the present work we use the third possibility for the
quantum Einstein-Cartan theory. It is well-known that the
pure quantum GR is renormalizable on-shell at the one-loop
level \cite{hove}. This enables one to consider, for instance,
the reduced on-shell version of renormalization group for the
Newton constant and cosmological term \cite{frts82}. Let us note
that this approach can be extended to become more informative when
the calculations are performed on the special background such as
deSitter space \cite{frts84}, but our intention here is to follow
a more simple method of \cite{frts82}. A very nice feature of this
approach is that the renormalization group equation for the
dimensionless combination of the cosmological and Newton
constants is gauge-fixing independent and, in this sense, is
well defined. Of course, the on-shell renormalization group
can not be seen as a completely consistent method, but it is
a useful starting point to deal with the QG theory.

The on-shell one loop renormalization group has been generalized
for the case of the Einstein-Cartan theory in \cite{BuSh-87}, in
the theory with an external axial vector current. It was shown
that the theory remains on-shell renormalizable at first loop in
the presence of such current and quantum torsion. Here we intend
to generalize these considerations in two ways, namely by including
an additional vector current and also by incorporating the Holst
term. We shall analyze to which extent the on-shell
renormalizability can be preserved in such a theory and also
consider the on-shell renormalization group to the extent it
is possible.

The paper is organized as follows. In Sect. 2, the classical
consideration of the Einstein-Cartan theory with the Holst term
and two (vector and axial vector) currents is presented. The
derivation of one-loop divergences and analysis of the on-shell
renormalizability of the theory is described in Sect. 3. Sect. 4
contains the consideration of the on-shell renormalization group
in the theory. Finally, in the last section we draw our
conclusions and discuss possible perspectives for a
future work.

\section{Simple representation for the Holst term}

In what follows we shall use the notations of
\cite{torsi},\footnote{One can use this reference and also many
other sources, e.g., \cite{Hehl-76,Gasperini,Hammond} for the
introduction to different aspects of gravity with torsion.} but
will first reproduce the main formulas, for the convenience of
the reader. The total action of gravity, including Einstein-Cartan
and Holst terms has the form
\beq
S_{EC} + S_H
&=&
- \,\frac{1}{\ka^2}\,\int d^4x\,\sqrt{-g}\,{\tilde R}
\,-\,\frac{1}{2\gamma\,\ka^2}\,\int d^4x\,\sqrt{-g}\,
\vp^{\al\be\mu\nu}\,{\tilde R}_{\al\be\mu\nu}\,,
\label{ECH}
\eeq
where $G=\ka^2/16\pi$ is Newton constant, also
$16\pi/\ka^2=M_P^2$. $\gamma$ is the Barbero-Immirzi parameter.
The scalar curvature is \
${\tilde R}=g^{\al\mu}g^{\be\nu}\,{\tilde R}_{\al\be\mu\nu}$
\ and $\,{\tilde R}_{\al\be\mu\nu}\,$ is the curvature tensor
depending on the  metric $g_{\al\be}$ and torsion
$T^\alpha_{\,\cdot\,\beta\gamma}$. This curvature is defined
on the basis of asymmetric connection
\beq
\tilde{\Gamma}^\alpha_{\,\beta\gamma}
- \tilde{\Gamma}^\alpha_{\,\gamma\beta}
= T^\alpha_{\,\cdot\,\beta\gamma} \neq 0\,.
\label{tor}
\eeq
Assuming that covariant derivative with torsion satisfies the
metricity condition $\tilde{\nabla}_\mu g_{\alpha\beta} = 0$, one
can easily derive the relation between affine connection
and Christoffel symbol ${\Gamma}^\alpha_{\;\beta\gamma}$,
\beq
\tilde{\Gamma}^\alpha_{\;\beta\gamma}
= {\Gamma}^\alpha_{\;\beta\gamma} + K^\alpha_{\cdot\;\beta\gamma}\,.
\label{con}
\eeq
Here the contorsion tensor is
\beq
K^\alpha_{\;\cdot\beta\gamma} = \frac{1}{2}
\left( T^\alpha_{\;\cdot\beta\gamma} -
T^{\;\alpha}_{\beta\cdot\gamma} -
T^{\;\alpha}_{\gamma\cdot\beta} \right)\,.
\label{1.7}
\eeq
The corresponding relations for curvature tensor and scalar with torsion
have the form
\beq
{\tilde{R}}^\la_{\cdot\,\tau\al\be} &=& R^\la_{\cdot\,\tau\al\be}
+ \na_\al\,K^\la_{\cdot\tau\be} - \na_\be\,K^\la_{\cdot\tau\al} +
K^\la_{\cdot\ga\al}\,K^\ga_{\cdot\tau\be}-
K^\la_{\cdot\ga\be}\,K^\ga_{\cdot\tau\al}\,,
\label{riemann}
\\
{\tilde {R}} &=& R + 2\,\na^\la\,K_{\cdot\,\la\tau}^\tau
-K_{\tau\la\,\cdot}^{\,\,\,\,\,\,\la}\,
K^{\tau\ga}_{\cdot\,\cdot\,\,\ga}
+ K_{\tau\ga\la}\,K^{\tau\la\ga}\,,
\label{R}
\eeq
where the quantities without tildes are Riemannian, without torsion.

One can introduce three irreducible components of torsion as follows:
\beq
\mbox{vector trace}
&&
T_{\beta} = T^\alpha_{\,\cdot\,\beta\alpha}\,,
\label{T}
\\
\mbox{axial vector trace}
&&
S^{\nu} = \epsilon^{\alpha\beta\mu\nu}T_{\alpha\beta\mu}\,,
\label{S}
\\
\mbox{tensor part}
&&
q^\alpha_{\,\cdot\,\beta\gamma}\,,
\label{q}
\eeq
when the last one satisfies the conditions
$\,q^\alpha_{\;\cdot\,\beta\alpha} = 0\,$
and $\,\epsilon^{\alpha\beta\mu\nu}q_{\alpha\beta\mu} =0$.

The generic torsion can be easily expressed as
\beq
T_{\alpha\beta\mu} = \frac{1}{3}
\left(T_{\beta}\,g_{\alpha\mu} - T_{\mu}\,g_{\alpha\beta}\right)
- \frac{1}{6}\, \varepsilon_{\alpha\beta\mu\nu}\,S^{\nu}
+ q_{\alpha\beta\mu}\,.
\label{irr}
\eeq
Now, replacing (\ref{irr}) into (\ref{1.7}) and (\ref{R}) we arrive at
\footnote{We correct a misprint in the coefficient of $T^2$ term
in \cite{torsi}.}
\beq
{\tilde R}
&=&
R - 2\,\na_\al T^\al - \frac{2}{3}\,
 T_\al \,T^\al
 + \frac{1}{2}\, q_{\alpha \beta \gamma}\,q^{\alpha \beta \gamma}
 +  \frac{1}{24}\, S_\al S^\al\,.
\label{Rtilde}
\eeq

Finally, repeating the same operation with (\ref{riemann}) and
then with the integrand of the Holst term, we arrive at the relation,
which was already reported in \cite{mercuri} (see also \cite{Hehl-2011}
for more detailed consideration and more complete list of references
on the history of the parity-violating terms in Einstein-Cartan theory,
\beq
\vp^{\al\be\mu\nu}\,{\tilde R}_{\al\be\mu\nu}
&=& -\,\na_\mu S^\mu\,-\,\frac23\,S^\mu\,T_\mu
\,+\, \frac12\,\vp^{\al\be\mu\nu}\,q^\la_{\,\cdot\,\al\be}\,q_{\la\mu\nu}
\nonumber
\\
&=&  -\,\na_\mu S^\mu \,-\,\frac23\,S \cdot T
\,+\,\frac12\,\vp\cdot q \cdot q\,.
\label{Holst}
\eeq
In the last relation we have introduced condensed notations with dots
for the contractions of two vectors and two tensors. The first of these
notations will be used a lot in what follows.

One can see that the first part of the Holst term is rather simple in
the representation
(\ref{Holst}). This term is nothing else but the simplest possible term
violating parity. In fact, this term was not introduced as a non-minimal
structure in the early works on quantum effects in gravity with torsion
\cite{bush1,BuSh-87} only because there was no interest to violate
parity. For instance, the non-minimal structure $\,\ph^2 S^\al T_\al\,$
becomes relevant in the scalar sector if the parity-breaking nonminimal
terms
$\,\bar{\psi}\ga^\al S_\al \psi\,$ or
$\,\bar{\psi}\ga^\al\ga^5 T_\al \psi\,$ are introduced.
In this case the Holst term can be easily obtained as part of the
induced action (extended Einstein-Cartan) of gravity with torsion,
e.g., it can result from some phase transition scheme, including
spontaneous symmetry breaking.

In order to better understand the effect of the Holst term,
let us include vector $V^\mu$ and axial vector $A^\mu$ fermion currents,
\beq
V^\mu &=& \eta_2\langle \bar{\psi} \ga^\mu \psi \rangle
\quad \mbox{and} \quad
A^\mu  \,=\, \eta_1\langle \bar{\psi} \ga^\mu \ga^5 \psi \rangle\,.
\label{currents}
\eeq
Let us note that the presence of non-minimal parameters $\eta_{1,2}$
is the condition of consistency of the theory at the quantum level,
especially if scalar fields and Yukawa interactions of these fields
with fermions are present \cite{bush1} (see also \cite{book} and
\cite{torsi} for extended discussions of this issue).
For the sake of compactness
of notations in the quantum part of the work, it is better to
introduce also rescaled currents \ $J^\mu = - \ka^2 A^\mu$ \ and
\ $W^\mu = - \ka^2 V^\mu$, such that the total action becomes
\beq
S_{t}
\nonumber
&=& S_{EC} + S_H
\,+\,\int d^4x\,\sqrt{-g}\,\big(V\cdot T + A\cdot S\big)
\label{total}
\\
\nonumber
&=& - \,\frac{1}{\ka^2}\,\int d^4x\,\sqrt{-g}\,
\Big\{
R + 2\La  - \frac{2}{3}\,T^2 + \frac{1}{2}\, q^2
+ \frac{1}{24}\,S^2
\\
&-&
\frac{1}{3\gamma}\,S \cdot T
- \frac{1}{2\ga}\,\na_\mu S^\mu
+ \frac{1}{4\ga}\,\vp\cdot q\cdot q + S \cdot J + T \cdot W \Big\}\,,
\eeq
where we also used compact notations \ $S^2 = S_\mu S^\mu$,
\ $T^2 = T_\mu T^\mu$ \ and \ $q^2 = q_{\mu\nu\tau}q^{\mu\nu\tau}$.

As usual in the Einstein-Cartan theory, torsion is not dynamical
field and can be integrated out. The dynamical equations for
different components of torsion have the form
\beq
- \frac{4}{3}\,T^\al - \frac{1}{3\gamma}\,S^\al + W^\al &=& 0\,,
\nonumber
\\
\frac{1}{12}\,S^\al - \frac{1}{3\gamma}\,T^\al + J^\al &=& 0\,,
\\
\nonumber
q^{\al\be\ga} &=& 0\,.
\label{3eqs}
\eeq
According to the last equation we will not consider the
component $q^{\al\be\ga}$ further. The first two equations
can be easily solved in the form
\beq \label{16}
T^\al
&=& \frac{3\gamma}{1+\gamma^2}\,
\Big(J^\al + \frac{\gamma}{4}\, W^\al\Big)\,,
\nonumber
\\
S^\al
&=& \frac{3\gamma}{1+\gamma^2}\,\Big(W^\al - 4\gamma J^\al\Big)\,.
\label{2sols}
\eeq
One can observe that the presence of parity-violating parameter
$\gamma$ leads to the mixing between vector and axial vector currents.
In principle, this mixing may have some strong phenomenological
consequences, and it would be interesting to explore its
consequences in particle physics. Such investigation could
lead to the upper bounds of certain combinations of the
Barbero-Immirzi parameter $\gamma$ and the non-minimal parameters
$\eta_{1,2}$, introduced in (\ref{currents}). However, in the
present work our purpose is not phenomenology, instead we shall
focus our attention on more formal aspects of the theory, related
to QG.

The dynamical equation for the metric in the theory (\ref{total})
leads to the on-shell relations
\beq
R_{\mu\nu} &=& D_{\mu\nu}
\,-\,
g_{\mu\nu}\,\Big(\La + \frac12\,S \cdot J + \frac12\,T \cdot W\Big)\,,
\label{Ricci}
\eeq
where we introduced a useful notation
\beq
D_{\mu\nu} &=& \frac23\,T_\mu T_\nu - \frac{1}{24}\,S_\mu S_\nu
+ \frac{1}{6\gamma}\,\big( S_\mu T_\nu + S_\nu T_\mu \big)
\quad
\mbox{and also} \quad D = D^{\mu}_{\ \mu}\ .
\label{Dmn}
\eeq
Finally, replacing (\ref{2sols}) into (\ref{Ricci}), after some
algebra we arrive at the on-shell relations
\beq
R_{\mu\nu}\Big|_{on-shell}  &=& \,-\,
\La g_{\mu\nu}\,+\,\frac{3\gamma}{1+\gamma^2}\,
\Big( 2\gamma J^2 - \frac{\gamma}{8}\,W^2 - J \cdot W\Big)g_{\mu\nu}
\nonumber
\\
&+& \frac{3\gamma}{1+\gamma^2}\,
\Big[\frac{\gamma}{8}\,W_\mu W_\nu  - 2\gamma J_\mu J_\nu
+ \frac{1}{2}\,\big(W_\mu J_\nu + W_\nu J_\mu\big)\Big]
\label{Ricci-on}
\eeq
and
\beq
R \Big|_{on-shell} &=& \,-\, 4\La \,+\,\frac{3\gamma}{1+\gamma^2}\,
\Big[ 6\gamma J^2 - \frac{3\gamma}{8}\,W^2
 - 3W\cdot J\Big]\,.
\label{R-on}
\eeq
Finally, for the total action (\ref{total}) on-shell we obtain
\beq
S_t\Big|_{on-shell} &=&
- \, \frac{1}{\ka^2} \int d^4x\,\sqrt{-g}\,
\Big\{\frac{3 \gamma}{1+\gamma^2}\,
\Big(4 \gamma J^2 - 2 J\cdot W - \frac{\gamma}{4}\, W^2\Big)
- 2 \La \Big\}\,.
\label{t-on}
\eeq
A simple observation concerning this action is as follows. In
the limit $\ga \to \infty$ the mixed term
with $(J\cdot W)$ goes to zero. This is of course a natural
feature, because this parity-violating term is only due to
the presence of the Holst term. This detail is an illustration
of the possible effects of the Holst term on the interaction
between the two vector currents.

\section{One-loop divergences off- and on-shell}

The divergences must be calculated on the basis of the off-shell
action (\ref{total}). We shall treat $g_{\mu\nu}$, $S_{\al}$ and $T_{\al}$
as quantum fields while $W^{\alpha}$ and $J^{\alpha}$ will be taken as
external sources. The Gaussian path integrals over $S_{\alpha}$ and
$T_{\alpha}$ do not generate divergences, because the corresponding
bilinear forms are $\,c$-number operators.
This means that integrations over these variables is greatly
simplified. Let us see this in more details.

Consider the background field method for the action (\ref{total})
and shift the field variable into background and quantum parts
according to
\beq
\label{22}
g_{\mu\nu} \rightarrow g'_{\mu\nu} = g_{\mu\nu} + \kappa h_{\mu\nu}
\,, \quad
S_{\mu} \rightarrow S'_{\mu} = S_{\mu} + \kappa \sigma_{\mu}
\,, \quad
T_{\mu} \rightarrow T'_{\mu} = T_{\mu} + \kappa t_{\mu}\,.
\eeq
The one-loop effective action depends on the bilinear in respect
to the quantum fields $h_{\mu\nu},\,\si_{\mu},\,t_{\mu}$ part of
the action. Since we are going to work with on-shell quantities,
the choice of the gauge fixing is irrelevant. For the sake of
simplicity we consider
\beq
S_{gf} = -\frac{1}{2\th}
\,\int d^4 x \sqrt{-g}\ \chi_{\mu}\ch^{\mu}\,,
\quad \mbox{where}\quad
\chi_\mu = \na_\la h_\mu^\la - \frac{\om}{2}\,\na_\mu h
\label{gauge}
\eeq
and chose the gauge fixing parameters in a way that leads to
the minimal bilinear form of the action, namely $\th=\om=1$.

The expansion performs as usual (see, e.g., \cite{book}
for details) and after some algebra we arrive at
\beq
\label{23}
S_{t}^{(2)} + S_{gf}
&=&
- \int d^4 x \sqrt{-g}\ \Bigg\{h^{\mu\nu}\Big[\,
\frac{1}{4}\Big(\delta_{\mu\nu,\al\be}
- \frac{1}{2} g_{\mu\nu} g_{\al\be}\Big)\Box
\,+\, \frac12\, R_{\mu\alpha\nu\beta}
\nonumber
\\
&+& \frac12\, g_{\nu\be}\, R_{\mu\al} \,-\,
\frac14\,\big(g_{\mu\nu} R_{\al\be} + g_{\al\be} R_{\mu\nu}\big)
\nonumber
\\
&-&
\frac14\Bigl(\de_{\mu\nu,\al\be} - \frac12\, g_{\mu\nu} g_{\al\be}\Bigl)
\Big( R + 2\La - \frac{2}{3}\,T^2 + \frac{1}{24}\,S^2
- \frac{1}{3\ga}\,S\cdot T + S\cdot J + T\cdot W\Big)
\nonumber
\\
&-&
\frac{1}{96}\,\big(g_{\mu\nu} S_\al S_\be
+ g_{\al\be} S_{\mu} S_{\nu}) - \frac{2}{3}\, g_{\mu\al} T_\nu T_\be
+ \frac{1}{6}\,(g_{\mu\nu} T_\al T_\be
+ g_{\al\be} T_\mu T_\nu\big)
\nonumber
\\
&-&
\frac{1}{6\ga} g_{\mu\al} (S_{\nu} T_{\be}
+ S_{\be} T_{\nu})
+ \frac{1}{12\ga} \big(g_{\mu\nu} S_\al T_\be
+ g_{\al\be} S_\mu T_\nu\big) +  \frac{1}{24}\, g_{\mu\al} S_\nu S_\be
\Big] h^{\al\be}
\nonumber
\\
&+&
\frac{1}{24}\, g^{\mu\nu} \sigma_{\mu} \sigma_{\nu}
- \frac{2}{3} g^{\mu\nu} t_{\mu} t_{\nu}
\,-\,\frac{1}{6\ga} \big(\si_\mu g^{\mu\nu} t_\nu
+ \si_\nu g^{\mu\nu} t_\mu\big)
 \nonumber
\\
&+&
h^{\mu\nu} \Big[ -\frac{1}{12}\, S_\mu \si_\nu
+ \frac{4}{3}\, T_{\mu} t_{\nu}
+ \frac{1}{3\ga}\, \big(S_\mu t_\nu + T_\mu \si_\nu)\Big]
\,+\,\Big[\frac{1}{24}\, g_{\al\be}\, S_{\mu} \si^\mu
\label{bi}
\\
&-&
\frac{2}{3}\, g_{\al\be} T_{\mu} t^\mu
\,-\, \frac{1}{6\ga}\, g_{\al\be} \,\big(T^\mu \si_\mu + S^\mu t_\mu\big)
+ \frac{1}{2}\, g_{\al\be} \,\si_\mu J^\mu
+ \frac{1}{2}\, g_{\al\be}\, t_{\mu} W^\mu\Big] h^{\alpha \beta}\Bigg\}\,,
\nonumber
\eeq
where $\,\de_{\mu\nu,\al\be} = (1/2)\,(g_{\mu\al} g_{\nu\be}
+ g_{\mu\be} g_{\nu\al})$. \
A relevant observation is that the path integral over $\si_\nu$ and
$t_\mu$ has the form
\beq
I &=&
\int dt_{\mu}\, d\si_\nu
\, \exp
\Bigg\{i \,\Big[\,\frac{1}{2}\,(\si_\mu\, t_{\mu}) (K^{\mu\nu})
{\si_\nu \choose t_{\nu}}
+ (\si_\mu\, t_\mu)\, {a^\mu \choose b^\mu} \Big] \Bigg\}\,,
\eeq
where $K^{\mu\nu}$ is a $c$-matrix and $a^\mu, b^\mu$ form a column depending on
the background fields. This non-derivative Gaussian integration gives
\beq
\label{24}
I &=& \exp \left\{
-\frac{i}{2} (a^\mu\ b^\mu) (K_{\mu\nu})^{-1}
 {a^\nu \choose b^\nu}\right\}\,.
\eeq
This is the same result which one could obtain just using the classical
equations of motion for the two components of torsion  $\si_\nu$ and
$t_\mu$. Since our intention is to calculate the on-shell effective
action, it means that we can simply ignore path integrals over
$\si_\nu$ and $t_\mu$. That means there is no need to perform the shift
of  $S_\nu$ and $T_\mu$ in (\ref{22}), instead one can directly use
corresponding classical equations of motion in the result of the
integration over quantum metric $h_{\mu\nu}$.

Finally, the relevant part of the bilinear expansion is
\beq
S_{t}^{(2)} + S_{gf} &=&
- \int d^4 x \sqrt{-g}\,\,
h^{\mu\nu}\,\Big\{\frac{1}{4}\Big(\de_{\mu\nu,\al\be}
- \frac12\,g_{\mu\nu} g_{\al\be}\Big) \Box
+ \frac{1}{2}\,R_{\mu\al\nu\be}
+ \frac{1}{2}\, g_{\nu\be}\, R_{\mu\al}
\nonumber
\\
&-& \frac{1}{4}(g_{\mu\nu} R_{\al\be} + g_{\al\be} R_{\mu\nu})
- \frac{1}{4}\Big(\delta_{\mu\nu,\al\be}
- \frac{1}{2}\,g_{\mu\nu} g_{\al\be}\Big)X
+ \frac{1}{4}\,Y_{\mu\nu,\alpha\beta}\Big\} \,\,h^{\alpha\beta}\,,
\label{26}
\eeq
where
\beq
\label{27}
X &=&
R + 2\Lambda - \frac{2}{3}\, T^2
+ \frac{1}{24}\,S^2
- \frac{1}{3\ga}\,S\cdot T + S\cdot J + T\cdot W
\eeq
and
\beq
Y_{\mu\nu,\alpha\beta}
&=&
\frac{1}{6}\, g_{\mu\alpha} S_{\nu} S_{\beta}
- \frac{1}{24}\,(g_{\mu\nu} S_{\alpha} S_{\beta}
+ g_{\alpha\beta} S_{\mu} S_{\nu})
- \frac{8}{3}\, g_{\mu\al}\, T_{\nu} T_{\be}
+ \frac{2}{3}\,(g_{\mu\nu}\, T_{\al} T_{\be}
+ g_{\al\be}\, T_{\mu} T_{\nu})
\nonumber
\\
&-& \frac{2}{3\ga}\, g_{\mu\alpha} (S_{\nu} T_{\beta} + S_{\beta} T_{\nu})
+ \frac{1}{3\ga} \,(g_{\mu\nu} S_{\alpha} T_{\beta}
+ g_{\alpha\beta} S_{\mu} T_{\nu})\,.
\label{28}
\eeq
Furthermore, the equation (\ref{26}) can be rewritten as
\beq
\label{29}
S_{t}^{(2)} + S_{gf} \nonumber &=&
- \int d^4 x \sqrt{-g}\ h^{\mu\nu}\Big(
\frac{1}{4}\,K_{\mu\nu,\al\be}\,\Box
+ \frac{1}{4}\,M_{\mu\nu,\al\be}\Big)\, h^{\alpha\beta}
\\
&=&
-\,\frac{1}{4}\,
\int d^4 x \sqrt{-g}\ h^{\mu\nu}\,K_{\mu\nu,\al\be}
\,\big[\de^{\al\be,}_{\quad \rho\sigma}\,\Box
+ \hat{\Pi}^{\al\be,}_{\quad \rho\si}]\, h^{\rho\sigma}\,,
\eeq
where
\beq
\nonumber
\label{33}
\hat{\Pi}_{\al\be,\rho\si}
&=&
2 R_{\rho\al\si\be} + 2 g_{\si\be}\, R_{\rho\alpha}
- \big(g_{\rho\si}\, R_{\al\be} + g_{\al\be}\, R_{\rho\si}\big)
\\
&+&
\frac{1}{2}\, g_{\rho\sigma} g_{\alpha\beta} R - \de_{\rho\si,\al\be} X
+ Y_{\rho\si, \al\be} - \frac{1}{2}\, g_{\rho\si}\,Y_{\rho\si, \al\be}\,g^{\mu\nu}\ \,,
\eeq
and
\beq
K^{-1}_{\mu\nu,\al\be} = K_{\mu\nu,\al\be} = \de_{\mu\nu,\al\be}
- \frac{1}{2} \,g_{\mu\nu} g_{\alpha\beta}\,.
\eeq

The one-loop contribution is given by the standard expression
\beq
\Ga^{(1)} &=& \frac{i}{2}\,\Tr \ln \big\{ \hat{K}\cdot
(\hat{\Box} + \hat{\Pi})\big\}
\,-\,i\,\Tr \ln \hat{H}_{ghost}\,.
\label{Ga}
\eeq
The ghost part does not depend on torsion or external currents, hence
the corresponding contribution will be identical to the standard one
for Einstein gravity \cite{hove}. Let us, therefore,
concentrate on the first term in (\ref{Ga}).
As far as
$\Tr \ln \hat{K} \,=\,\Tr \ln \hat{K}_{\mu\nu,\al\be}\,$
does not contribute to the divergences, they depend only on the matrix
$\,\hat{\Pi}^{\al\be,}_{\quad \rho\si}\,$
and also on the contribution of the Faddeev-Popov ghosts.

The practical calculation of divergences follows the standard scheme
\cite{hove} and we will avoid boring the reader with the details. The
result for the divergent part of the one-loop effective action can be
conveniently expressed via the tensor quantity (\ref{Dmn}) and has the
following final form:
\beq
\bar{\Ga}^{(1)}_{div}
&=&
- \,\frac{1}{\vp} \int d^4 x \sqrt{-g}\,
\Bigg\{\,\frac{53}{45}\, E + \frac{7}{10}\, R_{\mu\nu}^2
+ \frac{1}{60}\, R^2 + 8 D_{\mu\nu} D^{\mu\nu} - 2 D^2
\label{43}
\nonumber
\\
&+&
\frac{26}{3}\, R
\Big(\La + \frac{1}{2}\, S\cdot J + \frac{1}{2}\, T\cdot W\Big)
+ 20 \Big(\La + \frac{1}{2}\, S\cdot J + \frac{1}{2}\, T\cdot W\Big)^2
\Bigg\},
\eeq
where \ $E = R_{\mu\nu\alpha\beta}^2 - 4 R_{\mu\nu}^2 + R^2$ \
is the Lagrangian density of the Gauss-Bonnet term (Euler density).
Finally, $\vp=(4\pi)^2(n-4)$ is the parameter of dimensional
regularization. Let us remark that (\ref{43}) is relatively
simple due to some unexpected cancellations, for example of the
$\,DR$, $\,D_{\mu\nu}R^{\mu\nu}\,$ and a few other possible structures.

In order to formulate the on-shell renormalization group, we need
to use the classical equations of motion
(\ref{16}), (\ref{Dmn}), (\ref{Ricci-on}) and (\ref{R-on})
in eq. (\ref{43}). After some algebra we arrive at the result
\beq
\bar{\Gamma}^{(1)}_{div}\Big|_{on-shell}
&=&
-\, \frac{1}{\vp}\int d^4 x \sqrt{-g}\,\Bigg\{\,
\frac{53}{45}\, E - \frac{58}{5}\,\La^2
\,+\,\frac{81\,\ga^4 J^4}{(1 + \ga^2)^2}
+ \frac{81}{256}\,\frac{\ga^4 W^4}{(1 + \ga^2)^2}
\nonumber
\\
&+&
\frac{27}{40}\,\frac{\ga^2\,(43 \ga^2 + 58)}{(1 + \ga^2)^2}\, W^2\cdot J^2
+ \frac{241 \ga}{40(1 + \ga^2)}\,(16 \ga J^2 - \ga W^2
- 8 W\cdot J)\, \La
\nonumber
\\
&-& \frac{27\,\ga^2\,(W\cdot J)}{80\,(1 + \ga^2)^2}\,
 \Big[\big(56 +  116 \ga^2\big) (W\cdot J) + 240 \ga J^2 - 15 \ga W^2\Big]
\Bigg\}\,.
\label{44}
\eeq
An important difference between the expressions (\ref{43}) and (\ref{44})
is related to the gauge fixing dependence. The effective action (\ref{43})
has a lot of ambiguity related to the choice of the parameters $\th,\,\om$
in the action (\ref{gauge}). In fact, significant part of the terms can
be modified or even eliminated by an appropriate choice of these
parameters \cite{KTT}. On the other hand, there is no such gauge
dependence in the one-loop divergences for the on-shell effective
action \cite{frts82} (see also more detailed consideration in
\cite{a}), so the coefficients in the action (\ref{44}) do not
suffer from this ambiguity.

\section{On-shell renormalization group}

Our purpose is to construct the reduced on-shell version of the
Minimal Subtraction renormalization group. We shall use dimensional
regularization and hence it is necessary to formulate both
classical on-shell action (\ref{t-on}) and the on-shell
counterterm in $n$ space-time dimensions. The corresponding
expressions can be written in terms of the new notations
\beq
\widetilde{\la} &=& \al_1\la_1 + \al_2\la_2 + \al_3\la_3 + \al_4\la_4\,,
\label{ti-la}
\eeq
where
\beq
\label{80}
\la_1 = \ka^2 \La \,,
\qquad
\la_2 = \ka^2 J^2 \,,
\qquad
\la_3 = \ka^2 W^2 \,,
\qquad
\la_4 = \ga \,\ka^2(W\cdot J)
\eeq
on one side and
\beq
\nonumber
\widetilde{\sigma} &=&
\Om_{11}\la_1^2 + \Om_{22}\la_2^2 + \Om_{33}\la_3^2 + \Om_{44}\la_4^2
\label{si-tio}
\\
&+& \Om_{12}\la_1\la_2 + \Om_{13}\la_1\la_3 + \Om_{14}\la_1\la_4
+ \Om_{23}\la_2\la_3 + \Om_{24}\la_2\la_4 + \Om_{34}\la_3\la_4
\eeq
on another side.

The classical action and one-loop counterterms,
both on-shell (classical) have the form
\beq
\label{78}
S_t \Big|_{on-shell}
&=&
- \,\frac{1}{\ka^4} \int d^n x \sqrt{-g} \,\mu^{n-4}\,\widetilde{\la} \,,
\eeq
\beq
\label{79}
\De S^{(1)}\Big|_{on-shell}
&=&
\frac{1}{\vp}\,\cdot\,\frac{1}{\ka^4} \int d^n x \sqrt{-g} \,\mu^{n-4}\,
\widetilde{\si}\,.
\eeq
The coefficients in the expressions (\ref{ti-la}) and (\ref{si-tio})
can be taken directly from eqs. (\ref{t-on}) and (\ref{44})
\beq
\label{81}
\al_1 &=& -2, \qquad
\al_2 = \frac{12 \ga^2}{(1 + \ga^2)}\,,
\qquad
\al_3  = -\frac{3\ga^2}{4(1 + \ga^2)}\,,
\qquad
\al_4 = -\frac{6}{(1 + \ga^2)}
\eeq
and
\beq
\Om_{11}
&=&
-\frac{58}{5}, \,\quad \Om_{12} = \frac{482}{5}\frac{\ga^2}{(1 + \ga^2)} ,
\quad \Om_{13} = -\frac{241}{40}\frac{\ga^2}{(1 + \ga^2)}
, \quad
\Om_{14} = -\frac{241}{5(1 + \ga^2)}, \nonumber
\\
\quad \Om_{22} &=& \frac{81\ga^4}{(1 + \ga^2)^2},
\qquad
\Om_{23}
= \frac{27}{40}\frac{\ga^2(43\ga^2 + 58)}{(1 + \ga^2)^2}, \quad \Om_{24}
= -\frac{81\ga^2}{(1 + \ga^2)^2},
\label{Omegas}
\\
\Om_{33} &=& \frac{81}{256}\frac{\ga^4}{(1 + \ga^2)^2}, \quad \Om_{34}
= \frac{81}{16}\frac{\ga^2}{(1 + \ga^2)^2},
\quad
\Om_{44} = -\frac{(378 + 783\ga^2)}{20}\frac{1}{(1 + \ga^2)^2}\,.
\nonumber
\eeq

Let us note that consistent formulation of renormalization
group for both cosmological constant and Newton constant
(related to the inverse $\,\ka\,$ of the re-scaled Planck
mass) is definitely impossible since we are working in the
framework of the on-shell renormalization group. The form
of the classical action (\ref{78}) and the counterterms
(\ref{79}) indicate that there is no possibility to study
renormalization of $\ka$ in this framework, so in what
follows we will pursue only the aim of constructing the
renormalization group equations for effective charges
$\la_1$,  $\la_2$,  $\la_3$ and $\la_4$, defined in
(\ref{80}). One can also see this method as working in the
Planck units, where all quantities become dimensionless.

The on-shell renormalized action has the form which follows
from eqs. (\ref{78}) and (\ref{79}). Then the on-shell one-loop
divergences can be removed by means of renormalization
transformation
\beq
\widetilde{\la}_0 &=& \mu^{(n-4)}\Big(\widetilde{\la}
- \frac{\widetilde{\sigma}}{\vp}\Big) \,.
\label{reno}
\eeq
As far as $\widetilde{\la}_0$ does not depend on $\mu$, the
last relation implies that
\beq
\label{85}
(n-4)\,\Big(\widetilde{\la}
- \frac{\widetilde{\sigma}}{\vp}\Big)
\,+\, \Big(\mu \frac{d\widetilde{\la}}{d\mu}
\,-\, \frac{\mu}{\vp}\,\frac{d\widetilde{\sigma}}{d\mu}\Big) \,=\,0\,.
\eeq
Assuming that the divergent terms cancel, and using the
homogeneity property of $\widetilde{\sigma}$, we arrive at
the general $\be$-function for $\widetilde{\la}$ in $n$
space-time dimensions,
\beq
\label{88-n}
\be^n_{\widetilde{\la}} &=& -\,(n-4)\,\widetilde{\la}
\,-\,\frac{\widetilde{\sigma}}{(4\pi)^2} \,.
\eeq
Since our intention to to explore the renormalization group in
$n=4$, we have to take the limit $n \to 4$, to arrive at the
general renormalization group equation
\beq
\label{tota}
\frac{d\widetilde{\la}}{dt}
&=&
\mu\,\frac{d\widetilde{\la}}{d\mu}\,=\,
\be_{\widetilde{\la}} \,=\, -\,\frac{\widetilde{\sigma}}{(4\pi)^2} \,,
\eeq
where we introduced a useful parameter \ $t=\ln(\mu/\mu_0)$.

The next part of the work will be to extract the equations for
individual effective charges $\la_1$,  $\la_2$,  $\la_3$ and $\la_4$
from the single equation (\ref{tota}). This situation is definitely
more complicated than the one in the usual renormalizable theories,
and represents a necessary element of the more tricky scheme of
the on-shell renormalization group.

The case of the parameter  $\la_1$ has been considered in the
paper \cite{frts82}, where the on-shell  renormalization group
was invented. Let us suppose that the renormalization group
equation for the cosmological constant $\la_1$ does not depend
on the presence of external currents $J^\mu$ and $W^\mu$.
Setting $J^\mu=W^\mu=0$ we get $\la_{2,3,4}=0$ and then the
eq. (\ref{tota}) transforms into
\beq
\label{bla}
\al_1 \,\frac{d\la_1}{dt} &=& - \,\frac{1}{(4\pi)^2}\,\Om_{11}\,\la_1^2\,.
\eeq
Taking $\al_1$ and $\Om_{11}$ from (\ref{81}) and (\ref{Omegas}), one can
immediately obtain the corresponding equation of \cite{frts82},
\beq
\label{bla-fim}
\frac{d\la_1}{dt} &=& - \,\frac{29}{5\,(4\pi)^2}\,\la_1^2\,,
\eeq
indicating an asymptotic freedom for the dimensionless cosmological
constant in the UV for a positive cosmological constant and in the IR
for a negative cosmological constant.

One can follow similar approach for another effective
charge, $\la_2$. In this case one has to assume that when we
set $\La=0$ and $W^\mu=0$, the on-shell renormalization group
equation for the effective charge related only to $J^\mu$
does not change.  Then the considerations similar to the ones
which led us to (\ref{bla}) and (\ref{bla-fim}) provide us
with the equation
\beq
\label{88}
\frac{d\la_2}{dt}
&=&
\be_2\,=\,-\,\frac{\Om_{22}}{\al_2\,(4\pi)^2}\,\la_2^2 \,=\, -\,b_2^2\la_2^2\,
\,=\,-\,\frac{27}{4\,(4 \pi)^2}\,\frac{\ga^2}{(1 + \ga^2)}\,\la_2^2\,,
\eeq
indicating an asymptotic freedom for the dimensionless quantity
$\la_2$ in the UV, in case the vector $J^\mu$ is time-like, and
in the IR in case the same vector is space-like.

In a similar way one can obtain the equation for
the third parameter
\beq
\label{91}
\frac{d\la_3}{dt}
&=&
\be_3\,=\,-\,\frac{\Om_{33}}{\al_3\,(4\pi)^2}\,\la_3^2 \,=\,b_3^2\la_3^2
\,=\,\frac{27}{64\,(4 \pi)^2}\,\frac{\ga^2}{(1 + \ga^2)}\,\la_3^2\,.
\eeq
In this case we observe the asymptotic freedom for the dimensionless
quantity $\la_3$ in the UV if vector $W^\mu$ is space-like, and
in the IR in case this vector is time-like.  For the sake of
simplicity, we shall assume that the initial value of
$\la_{2}(\mu_0)=\la_{2}^0$ is positive and that the initial
value $\la_{3}(\mu_0)=\la_{3}^0$ is negative. In this case we
have asymptotic freedom for both charges in UV and will try to
explore this limit in what follows. It is important to note that
the signs of $\la_2$ or $\la_3$ are not limited by the arguments
of stability or alike, in particular because they correspond to the
properties of external (non-dynamical) currents.

Now we can start solving a more complicated problem of
formulating the on-shell renormalization group equation for the
effective charge $\la_4(t)$ and eventually for the effective
Barbero-Immirzi parameter $\ga(t)$.
By subtracting eqs. (\ref{88}) and (\ref{91}) with the factors
 $\al_2$ and $\al_3$, from eq. (\ref{tota}) we obtain
\beq
\al_4 \,\frac{d\la_4}{dt}
&=&
\frac{1}{(4\pi)^2}\big(
- \,\widetilde{\si}\,+\,\Om_{22}\la_2^2\,+\,\Om_{33} \la_3^2\big)
\nonumber
\eeq
that directly brings us to
\beq
\label{92}
(4\pi)^2\,\frac{d\la_4}{dt}
&=&
-\,\frac{1}{\al_4}\,\big(
\Om_{44}\la_4^2 + \Om_{23}\la_2\la_3 + \Om_{24}\la_2\la_4
+ \Om_{34}\la_3\la_4\big)\,.
\eeq
This is the renormalization group equation for the running parameter
$\la_4(t)$. The $\be$-function here depends on $\la_2(t)$ and
$\la_3(t)$, so the first impression is that one can solve eq.
(\ref{92}) only after solving eqs. (\ref{88}) and (\ref{91}).
However, the real situation is much more complicated. The parameter
$\,\la_4\,$ is strongly related to $\,\la_2\,$ and
$\,\la_3$, because all three constants are constructed from
two fermion currents, $J^\al$ and $W^\al$, via eq. (\ref{80}). In fact,
we have found the room for an independent equation (\ref{92}) only
because $\,\la_4\,$ depends not just on the magnitude of the
currents $J^\al$ and $W^\al$, but also on the angle between them
and on the Barbero-Immirzi parameter $\ga$. In what follows we
assume that the mentioned angle does not run with the scale. This
feature enables one to construct the renormalization group
equation for $\ga$.

Let us derive the renormalization group equation for the
Barbero-Immirzi parameter. For this end we return to the equations
(\ref{88}) and (\ref{91}). Since $\,\ka\,$ is a universal constant
(inverse Planck mass), one has to assume that the external currents
themselves are running quantities, that means
$J^\al = J^\al(\mu)$ and $W^\al = W^\al(\mu)$.
By using (\ref{80}), one can rewrite (\ref{88}) and (\ref{91}) as
\beq
\label{93}
\frac{d\la_2}{dt}
&=& \ka^2 \ \frac{dJ^2}{dt} = 2\ka^2 J^\al \ \frac{d J_\al}{dt}
= -\frac{\Om_{22} \la_2^2}{\al_2 (4\pi)^2} = \be_2 \,,
\\
\label{94}
\frac{d\la_3}{dt}
&=& \ka^2 \ \frac{dW^2}{dt} = 2\ka^2 W^\al \ \frac{d W_\al}{dt}
= -\frac{\Om_{33} \la_3^2}{\al_3 (4\pi)^2} = \be_3 \,.
\eeq
Let us make a natural assumption that
\beq
\label{102}
\frac{d J_\al}{dt} \,=\, \Theta_2 J_\al\,.
\eeq
Than it is easy to show that the
\beq
\Theta_2 &=& -\,\frac{\Om_{22} \ \ka^2 J^2}{2 \al_2 (4\pi)^2} \,.
\nonumber
\eeq
In the same way, we find
\beq
\label{103}
\frac{d W_\al}{dt} \,=\, \Theta_3 W_\al\,,\qquad
\mbox{where} \qquad
\Theta_3 \,=\, -\,\frac{\Om_{33} \,\ka^2 W^2}{2 \al_3 (4\pi)^2} \,.
\eeq
Then we have two relations,
\beq
\label{104}
\frac{d J_\al}{dt} = -\frac{\Om_{22} \ \ka^2 J^2}{2 \al_2 (4\pi)^2}\,J_\al
\qquad \mbox{and}  \qquad
\frac{d W_\al}{dt} = -\frac{\Om_{33} \ \ka^2 W^2}{2 \al_3 (4\pi)^2}\, W_\al\,.
\eeq
As far as $\,\la_4 = \ga \ka^2\, (W\cdot J)$,
the renormalization group equation for $\la_4$ is a consequence of
equations (\ref{104}) and the running of $\ga$, which we also
want to find. In this way one can obtain
\beq
\label{105}
\frac{d\la_4}{dt}
&=&
\ka^2 \,\Big(\,\frac{d\ga}{dt}\,
W\cdot J \,+\, \ga\, J^\al \,\frac{dW_\al}{dt}
\,+\, \ga\, W^\al\, \frac{d J_\al}{dt}\,\Big)\,.
\eeq
Replacing (\ref{92}) into (\ref{105}) and using eqs. (\ref{104}), we arrive at
\beq
\label{106}
(4\pi)^2\,\frac{1}{\ga}\, \frac{d\ga}{dt}
&=& -\frac{\Om_{44} \la_4}{\al_4}
\,+\, \la_2\,
\Big(\frac{\Om_{{22}}}{2 \al_2}-\frac{\Om_{24}}{\al_4}\Big)
\,+\,
\la_3\,
\Big(\frac{\Om_{{33}}}{2 \al_3}-\frac{\Om_{34}}{\al_4}\Big)
\, - \, \frac{\la_2 \la_3}{\al_4\la_4}\,\Om_{23}\,.
\eeq
The last equation describes the renormalization group running of
the Barbero-Immirzi parameter within the on-shell renormalization
group scheme. In this consideration we assumed that the angle
between the four-dimensional currents $J^\mu$ and $W^\mu$ does
not run with the renormalization group scale. This is a small
price to pay for the possibility to consider renormalization
group in the non-renormalizable theory such as Einstein-Cartan
gravity with the Holst term.

The problem of exploring the asymptotic behavior of the effective
charges $\la_{2,3,4}(t)$ and $\ga(t)$ on the basis of eqs.
(\ref{88}), (\ref{91}), (\ref{92}) and (\ref{106}) turns out to be
very complicated, and unfortunately we were unable to solve it
in a completely satisfactory way. Let us present only some part
of consideration, which can be useful to show what is the origin
of the difficulties.

The simplest assumption is that all four parameters
$\,\la_{2,3,4}(t)\,$ and $\,\ga(t)\,$ have moderate running
and therefore one can work in the leading-log approximation.
Then the eqs. (\ref{88}) and (\ref{91}) can be easily solved
for a constant $\,\ga\,$ and give
\beq
\label{spec}
\la_2(t) &=& \frac{\la_{20}}{1+b_2^2\la_{20}\,t}
\,,\qquad
\la_3(t) \,=\, \frac{\la_{30}}{1-b_3^2\la_{30}\,t}\,.
\eeq
In this case equation (\ref{92}) can be easily cast into
the form
\beq
\frac{d\la_4}{dt}
&=&
A(t) \la_4^2 + B(t)\la_4 + C(t)\,.
\label{ABC}
\eeq
Mathematically, (\ref{ABC}) is a Riccati equation, which can be
solved if we first get some particular solution. In order to
achieve this, we can make some simplifications.
Consider an asymptotic regime, assuming
$\,(\be_2 /\la_2^0)t \gg 1\,$
and $\,(\be_3 /\la_3^0)t \gg 1$, such that, approximately,
\beq
\la_{2,3}(t) \,=\,\frac{l_{2,3}}{t}\,,
\quad
\mbox{where}
\quad
l_2\,=\,\frac{4\,(4\pi)^2(1+\ga^2)}{27\,\ga^2}
\quad
\mbox{and}
\quad
l_3\,=\,16 l_2\,.
\label{sol23}
\eeq

In this way eq. (\ref{ABC}) becomes simpler,
\beq
\frac{d\la_4}{dt}
&=&
A_0 \la_4^2 + \frac{B_0}{t}\,\la_4 + \frac{C_0}{t^2}\,,
\label{ABC-0}
\eeq
where
\beq
A_0 &=& -\frac{\Om_{44}}{\al_4\,(4\pi)^2}
\,,\quad
B_0 \,=\, -\frac{\Om_{24}\,l_2+\Om_{34}\,l_3}{\al_4\,(4\pi)^2}
\,,\quad
C_0 \,=\, -\frac{\Om_{23}\,l_2\,l_3}{\al_4\,(4\pi)^2}
\,.
\label{ABC-0-coefs}
\eeq
It is quite natural to look for a particular solution
of eq. (\ref{ABC-0}) in the form
\beq
\la_4(t) &=& \frac{l_4}{t}\,,
\quad \mbox{where} \quad
l_4 = -\,\frac{B_0+1}{2\,A_0}
\pm \frac{1}{2\,A_0}\sqrt{(B_0+1)^2-4C_0 A_0}\,.
\label{sol}
\eeq
In case of a real {\it r.h.s.} of the last expression, one
can easily show that an arbitrary particular solution is
asymptotically approaching (\ref{sol}).
Unfortunately, a direct calculus shows that the root in
eq. (\ref{sol}) has only solutions with non-zero imaginary
part. This feature leaves very small chances to find a fixed
point for the system of equations (\ref{88}), (\ref{91}),
(\ref{92}) and (\ref{106}). According to (\ref{80}), the
parameter $\,\la_4\,$ can be complex only due to the complex
parameter $\,\ga$. This means that the ratio between real
and imaginary parts of $\,\la_4\,$ and $\,\ga\,$ should be
identically equal. However, direct calculations show that
this situation contradicts the equations (\ref{92}) and
(\ref{106}). This means that the system of renormalization
group equations for the effective parameters $\,\la_{2,3,4}(t)\,$
and $\,\ga(t)\,$ has no fixed points.

The absence of the Holst term means the limit $\,\ga \to \infty$
for the Barbero-Immirzi parameter. An inspection of the eqs.
(\ref{88}),\ (\ref{91}) and (\ref{92}) with $\Om_{22}$ and $\Om_{33}$
defined in (\ref{Omegas}) shows that in this limit there are
usual UV fixed points, which can correspond to the asymptotic
freedom in the parameters  $\,\la_2(t),\,\la_3(t)$ under the
right choice of initial conditions. Therefore, the role of the
Holst term in the renormalization group is very strong. Our
results show that the presence of finite $\ga$ breaks down the
simple form of the renormalization group flows and leads to much
more complicated scale behavior which looks irregular, at least
at the present stage of investigating the problem.

In this situation a natural question to ask is whether the
limit $\,\ga \to \infty$ for the Barbero-Immirzi parameter is
smooth. It is easy to see that in this limit we also have
$\,\la_4 \to \infty$. Therefore the smooth limit concerns
the ratio between the two effective parameters,
$\,p= \la_4/\ga$\,. \
The equation for this ratio can be easily obtained from eqs.
(\ref{80}), (\ref{102}) and (\ref{103}). After a very small
calculus we arrive at the equation
\beq
\frac{dp}{dt} &=&
-\, \frac{p}{2\,(4\pi)^2}\,
\Bigg[ \frac{\Om_{22}}{\al_2}\, \la_2(t)
\,+\,
\frac{\Om_{33}}{\al_3}\, \la_3(t)\Bigg]\,, \qquad
p(0) \,=\, p_0\,.
\label{p}
\eeq
Using the asymptotic estimates for $\,\ga \to \infty$,
\beq
\Om_{22} \propto 81\,,\qquad
\Om_{33} \propto \frac{81}{256}\,,\qquad
\al_2 \propto 12\,,\qquad
\al_3 \propto -\frac34\,,
\nonumber
\eeq
we arrive at the solution of (\ref{p}),
\beq
\frac{p(t)}{p_0} & \propto &
\Big(1+b_2^2\la_{20}\,t\Big)^{-1/2}\,\,
\Big(1-b_3^2\la_{30}\,t\Big)^{-1/2}\,.
\label{p-sol}
\eeq
The last formula shows that we can switch off the Barbero-Immirzi
parameter smoothly and the ratio $\,p(t)\to 0\,$ asymptotically at
$\,t \to \infty\,$ in the same way as the effective charges
$\,\la_2(t)\,$ and  $\,\la_3(t)$. This shows that our hypothesis
of a non-running angle between two currents is correct in the
regime of very small Holst term, at least. However, this
confirmation concerns only this special limit.

Of course, the most interesting part is the running for a
finite Barbero-Immirzi parameter, but in this case we could
not achieve a reliable analytic estimate of the results.
In this situation one can rely only on the numerical solution
for the system of equations  (\ref{88}), (\ref{91}), (\ref{92})
and (\ref{106}). The corresponding analysis has been done,
however the output shows very strong dependence on the choice
of initial conditions and after all, there is no convincing
qualitative interpretation of the results. For this reason, we
decided not to bother the reader with the technical details here.
One could imagine that the situation may become different in a more
complete case when we also take the running of $\la_1$ into
account. The technically more cumbersome analysis of this case
have been performed and we saw that there are no much changes.
Qualitatively, the situation remains
the same, that is there are no nontrivial fixed points in
the presence of finite Barbero-Immirzi parameter.

\section{Conclusions}

We have considered the Einstein-Cartan theory with an additional
Holst term, which plays an important role in loop quantum gravity
\cite{PR,freidel}. In classical theory this term is
well-known to identically vanish for zero torsion, it manifest
itself only in the presence of fermion currents. Following
\cite{mercuri}, we used the irreducible components of torsion to
write the Holst term in a simple form, where its parity-violating
nature becomes clear.

In the main part of the paper we performed one-loop calculations
in the Einstein-Cartan theory with the Holst term, cosmological
constant and two external fermion currents, namely with vector and
axial vector ones. As one should expect, the divergences do not
repeat the form of the classical action. On the other hand, the
divergences have strong gauge-fixing dependence. In pure quantum
GR one can chose the gauge-fixing in such a way that the one-loop
$S$-matrix is becoming finite \cite{KTT}, however this in not the
case if the matter is present, including fermions. Indeed, one
does not need to calculate explicitly gauge-fixing dependence,
it is sufficient
to remember that, at the one-loop level, this dependence disappears
on the classical mass-shell in a general gauge theory \cite{LVT} (see
also \cite{Lavrov-FRG} for a recent review of the subject).

The real problem is how to extract the potentially relevant
physical information from the gauge-dependent effective action.
One of the simplest possibilities has been suggested by Fradkin
and Tseytlin in \cite{frts82}, where the truncated, on-shell,
version of renormalization group equations has been introduced.
Within this scheme one can arrive at the gauge-invariant form
of running for the dimensionless combination of the cosmological
and Newton constants. The on-shell renormalization group has
been also used in the Einstein-Cartan theory with axial vector
current \cite{BuSh-87}, but the situation becomes much more
complicated and interesting in the presence of the Holst term.

In is clear that the on-shell renormalization group equations have
much more restricted theoretical background than the conventional
renormalization group in renormalizable theories. However, even in
the non-renormalizable theory such as Einstein-Cartan with the Holst
term we were able to establish the renormalization group equations
for all dimensionless effective charges, including cosmological
constant, squares of both fermionic currents, their mixing and
finally, for the Barbero-Immirzi parameter $\,\ga$.
Unfortunately, the equations which we have obtained are very
complicated and do not enable us to apply standard treatments.
In particular, we were unable to find non-trivial UV fixed points
in the theory or establish, by means of numerical methods,
some reliable form of the renormalization group trajectories
for the dimensionless effective charges.

Finally, let us present a short discussion of the perspectives
to extend our results. The set of equations which we have obtained
here, can be seen as a low-energy approximation for the
renormalization group in the theory with full UV completion,
which is supposed to be renormalizable. In the present case such
a complete theory should include higher derivatives in the metric
sector \cite{Stelle-77} and kinetic terms for torsion (see the
discussion in \cite{book,torsi}). Only quantum calculations in
such a full theory coupled to fermions \cite{BKSVW,Yukawa-ia,book}
can provide a completely reliable form of the renormalization group
equations in the theory with Barbero-Immirzi parameter. In practise,
the derivation of such equations is possible but promise to be very
involved, so we leave it for the possible future work. At the same
time, certain technical tools which we developed here will be
certainly necessary for such a calculation.

\section*{Acknowledgements}
Authors are grateful to Guilherme de Berredo-Peixoto and Cleber Abrahão
de Souza for useful discussions about the Holst term. We are especially
grateful to Prof. Friedrich Hehl, who explained us a mistake in the Eq.
(\ref{Holst}) in the first version of the manuscript and also provided us
by some relevant references. The work of the authors has been partially
supported by CAPES, CNPq, FAPEMIG and ICTP (I.Sh.).



\end{document}